\documentclass{jfm}
\usepackage{graphicx}
\newcommand{\uptau}{\tau}

\usepackage{natbib}
\usepackage[justification=justified]{caption}
\usepackage{amsmath,bm}

\usepackage[colorlinks=true,linkcolor=blue,citecolor=blue,urlcolor   = blue]{hyperref}
\usepackage[nameinlink]{cleveref}
\crefname{equation}{}{}
\crefname{eqnarray}{}{}
\crefname{figure}{figure}{figures}
\Crefname{Figure}{Figure}{Figures}
\crefname{table}{table}{tables}

\usepackage{amsfonts}
\usepackage{booktabs}

\usepackage{color}
\usepackage{soul} 

\definecolor{darkgreen}{rgb}{0,0.5,0.0}
\definecolor{purple}{rgb}{0.7,0.0,0.5}

\newcommand{\RomanNumeralCaps}[1]

\title{Floating active carpets drive transport and aggregation in aquatic ecosystems}

\author{G. Aguayo\aff{1},
A. J. T. M. Mathijssen\aff{2},
\and
H. N. Ulloa\aff{3},
R. Soto\aff{1} and
F.~Guzm{\'a}n-Lastra\aff{4}\corresp{\email{fguzman@uchile.cl}}}

\affiliation{
\aff{1} Departamento de F\'isica, Facultad de Ciencias F\'isicas y Matem\'aticas, Universidad de Chile, Beauchef 850, Santiago Chile
\aff{2} Department of Physics and Astronomy, University of Pennsylvania, Philadelphia, USA
\aff{3} Department of Earth and Environmental Science, University of Pennsylvania, Philadelphia, USA

\aff{4} Departamento de F\'isica, Facultad de Ciencias, Universidad de Chile, Las Palmeras 3425, Santiago Chile.}

\begin{document}
\maketitle

\begin{abstract} 
Communities of swimming microorganisms often thrive near liquid-air interfaces. 
We study how such `active carpets' shape their aquatic environment by driving biogenic transport in the water column beneath them. The hydrodynamic stirring that active carpets generate leads to diffusive upward fluxes of nutrients from deeper water layers, and downward fluxes of oxygen and carbon. 
Combining analytical theory and simulations, we examine the biogenic transport by studying fundamental metrics, including the single and pair diffusivity, the first passage time for particle pair encounters, and the rate of particle aggregation.
Our findings reveal that the hydrodynamic fluctuations driven by active carpets have a region of influence that reaches orders of magnitude further in distance than the size of the organisms. 
These nonequilibrium fluctuations lead to a strongly enhanced diffusion of particles, which is anisotropic and space-dependent.
Fluctuations also facilitate encounters of particle pairs, which we quantify by analysing their velocity pair correlation functions as a function of distance between the particles.
We found that the size of the particles plays a crucial role in their encounter rates, with larger particles situated near the active carpet being more favourable for aggregation. 
Overall, this research broadens our comprehension of aquatic systems out of equilibrium and how biologically driven fluctuations contribute to the transport of fundamental elements in biogeochemical cycles.
\end{abstract}

\vspace{-1.5cm}
\section{Introduction} 
\label{sec:Introduction}

Swimming microorganisms that accumulate into dense floating films are widespread at interfaces of aquatic environments \citep{zhan2014accumulation,sengupta2017phytoplankton, durham2012thin, mathijssen2016hydrodynamics, desai2020biofilms, vaccari2017films,sepulveda2021persistence}. The emergence of these `\textit{active carpets}' \citep{mathijssen2018nutrient, guzman2021active} (AC or ACs for plural) results from various factors, including hydrodynamic interactions, mechanical responses, taxis and the optimisation of metabolic activities, to name a few \citep{berke2008hydrodynamic,durham2012thin, sommer2017bacteria, desai2020biofilms, ahmadzadegan2019hydrodynamic}. Recently, \citet{mathijssen2018nutrient} introduced a framework to quantify how ACs generate hydrodynamic fluctuations in their surrounding fluid. Depending on the swimming strategy of the microbes, ACs can either attract or repel fluid in the presence of orientational and/or density gradients \citep{mathijssen2018nutrient, kanale2022spontaneous, lambert2013active}, and generate anisotropic material transport in the water column \citep{guzman2021active}. 
Previous studies have shown that ACs can facilitate nutrient transport and enhance diffusion orders of magnitude larger than thermal fluctuations. However, it remains unexplored how such ACs can mobilise particles to and from liquid-air interfaces and impact the clustering dynamics of suspended matter within aquatic systems.

The active material transport induced by living and artificial microswimmers stands at the forefront of fluid dynamics research \citep{angelani2011effective,gokhale2022dynamic,madden2022hydrodynamically,omar2018swimming, dani2022hydrodynamic}. It encompasses diverse phenomena, starting with active diffusion in which tracer particles are entrained due to hydrodynamic flows induced by swimming microorganisms \citep{mathijssen2018universal, jeanneret2016entrainment, pellicciotta2020entrainment, jin2021collective,bardfalvy2024collective,vskultety2024hydrodynamic}, generating an effective enhanced diffusion over the summation of these encounters. This active diffusion has been measured experimentally for various microorganisms and via numerical simulations, predicting a fundamental relationship between the active flux, the persistence of swimmers, and the relative distance between microswimmers \citep{morozov2014enhanced, pushkin2013fluid,de2017stirring, mino2011enhanced,mino_dunstan_rousselet_clement_soto_2013}. 

Although particles disperse due to active diffusion, their final concentration profile is restricted because of fluid confinement \citep{hamada2020diffusion,morozov2014enhanced}. In the case of ACs settling near non-slip boundaries, we find non-Boltzmannian concentration distributions \citep{guzman2021active}. This counterintuitive behaviour suggests the existence of an active temperature gradient, which has not been measured before in the context of biogenic hydrodynamic diffusion \citep{loi2008effective, takatori2015towards, ortlieb2019statistics}. 

When a microorganism swims, it induces local flows that, when superimposed over an entire colony of microswimmers, may generate coherent flows and large-amplitude disturbances. Examples of the impact of collective motions on scalar transport at intermediate Reynolds numbers, $Re\sim O(1)$, include the biogenic mixing induced by zooplankton diel vertical migration and bioconvection observed in natural aquatic systems and laboratory experiments \citep{wang2015biogenic, dabiri2010role, simoncelli2017can, javadi2020photo, pedley1992bioconvection, hill2005bioconvection, sommer2017bacteria,noto2023simple}. Yet, biomixing owing to active turbulence \citep{alert2022active} in ACs remain unexplored. 

A rigorous quantification of biogenic mixing requires the analysis of tracer pair dispersion, rather than single particle diffusivity \citep{belan2019pair}. Pair dispersion provides information not only about the mixing capacity of a system but also about particle aggregation. The latter is a crucial mechanism for sustaining aquatic life and a fundamental skill of biological cells \citep{burd2009particle,cruz2022particle, font2019collective, arguedas2022elongation, camassa2019first,maheshwari2019colloidal}. Recent studies have deepened our insights into the dynamic clustering and aggregation of colloids \citep{jia2019reversible, angelani2011effective,dani2022hydrodynamic}. Notably, these studies have revealed the role of colloid size in governing pairwise interactions with microswimmers \citep{kushwaha2023phase,gokhale2022dynamic}. Factors such as tumbling rate and chirality have been found to influence short-range encounters, diffusion rates, and mixing \citep{madden2022hydrodynamically, belan2019pair}. However, our current understanding of the hydrodynamic interactions induced by microswimmers living near fluid-fluid interfaces on mixing and colloid clustering remains limited \citep{gonzalez2021impact,dani2022hydrodynamic, wang1998collision}. 

In this paper, we build upon the singularity method to delve into the dynamics of ACs living near fluid-air interfaces and their impact on tracer dynamics. Our objectives are twofold. Firstly, we seek to expand the theory of ACs by providing analytical solutions for the velocity fluctuations induced by a colony of microswimmers inhabiting fluid-air interfaces. Secondly, we investigate suspended tracer dynamics, quantifying universal metrics, including the single and pair diffusivities, and particle aggregation driven by ACs. We start by presenting an overview of the theoretical framework for ACs and the numerical methods applied to analyse both ACs and tracer dynamics. Finally, we report our findings and draw conclusions.

\section{Active carpets}
\label{sec:ActiveCarpets}
\begin{figure}
\centerline{\includegraphics[width=1\linewidth]{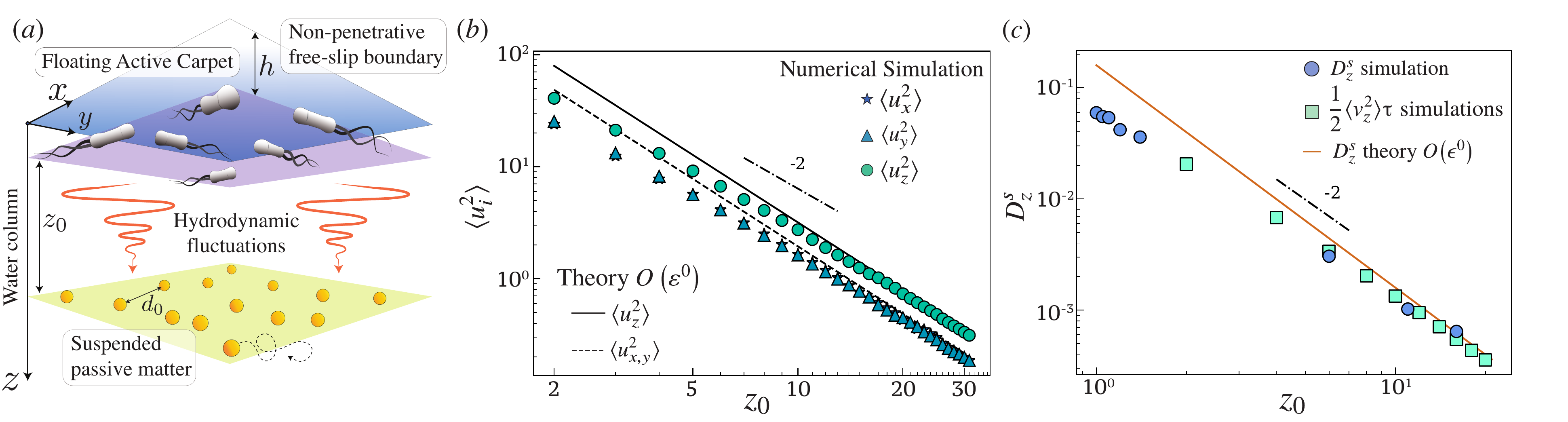}}
\captionsetup{width=1\linewidth}
\caption{
($a$) An active carpet near the liquid-air interface generates flows that stir particles suspended below. ($b$) Strength of hydrodynamic fluctuations as a function of distance from the AC. Markers represent simulation results and lines represent the theoretical prediction from equation~\eqref{variances_b}. ($c$) Single-particle diffusivity in the $z$ direction. Circles are results from simulated MSDs, the squares are results from simulating the variance following equation~\eqref{FD_MSD}, and the solid orange line represents the theoretical prediction from equations~\eqref{variances_b} and \eqref{FD_MSD}.}
\label{fig:flowfield}
\end{figure}

We consider microswimmers in a 3D semi-infinite fluid that self-organise into an AC, as illustrated in figure~\ref{fig:flowfield}($a$).
The fluid-air interface is located at $z=0$, which satisfies non-penetrative and free-slip boundary conditions.
The AC is situated just beneath the interface, at a distance $z=h$, where $h$ is the typical swimmer size.
We use a Cartesian coordinate system where the positive $z$ direction points downward from the liquid-air interface so that $z$ indicates depth.
To analyse the flow generated by each swimmer, we utilise a multipole expansion of the fundamental solution of the Stokes equations together with the method of images \cite[see e.g.][]{lauga2020fluid}. 
For a swimmer located at $\bm{r}_s$, an image swimmer is located at the mirror position $\bm{r}_s^*=\mathsf{M}\cdot \bm{r}_s$, where $\mathsf{M}$ represents the mirror matrix $\mathsf{M}=\text{diag}(1,1,-1)$. Microswimmers, such as motile bacteria, can be considered as force dipoles in the far-field approximation, as shown by \citet{drescher2011fluid}. Defining the orientation of the swimmer $\bm{\hat{p}}$ as the unit vector along the axis that connects the flagella with the head of the swimmer, the fluid flow at a position $\bm{r}$ caused by a swimmer located at position $\bm{r}_s$ is \citep{mathijssen2016hydrodynamics}
\begin{equation}
 \label{totstresslet}
   \bm{u}_s(\bm{r},\bm{r}_s,\bm{\hat{p}})=  -\kappa\hat{\bm{p}}\cdot\nabla_{s}\bm{v}-\kappa\hat{\bm{p}}\cdot\nabla^*_{s}\bm{v}^*,
\end{equation}
where $\bm{v} = \mathsf{G}(\bm{r}-\bm{r}_s)\cdot \hat{\bm{p}}$ is the Stokeslet flow and  $\bm{v}^{*} = \mathsf{M}\,\mathsf{G}(\bm{r}-\bm{r}_s^*)\cdot \hat{\bm{p}}$ is its image,  with $\mathsf{G}(\bm{x})=\frac{1}{8\pi |\bm{x}|}(\mathsf{I}+\bm{x} \bm{x}/|\bm{x}|^2)$ the Oseen tensor.  
The coefficient $\kappa$ is the dipole strength that can be expressed in terms of the exerted force by the swimmer in the fluid, its characteristic length, and the fluid viscosity \citep[see e.g.][]{mathijssen2016hydrodynamics}. In general, the dipole strength sign depends on the swimming strategy and the swimmer body-geometry: $\kappa > 0$ represents a \textit{pusher}, whereas $\kappa < 0$ represents a \textit{puller} \citep{happel1983low}.

We consider an AC composed of point dipoles, each of them located at positions $\bm{r}_s(t)=(x_s(t),y_s(t),h)$ and oriented along $\bm{\hat{p}}(t)=(p_x(t),p_y(t),0)$, forming a flat monolayer of microswimmers as is sketched in figure~\ref{fig:flowfield}($a$). At any time $t$, microswimmers exert a flow field given by equation~\eqref{totstresslet} so the total flow field generated by an AC is given by $\bm{u}(\bm{r},t)=\sum_s\bm{u}_s(\bm{r},\bm{r}_s(t),\hat{\bm{p}}(t))$ 

with, 
\begin{equation}
\begin{split}
\bm{u}_s\left(\bm{r},\bm{r}_s,\bm{\hat{p}}\right)=\kappa \Bigg[ &(x-x_s) \left(\left(\frac{-1}{r_{-}^{3/2}}+\frac{-1}{r_{+}^{3/2}} \right)+ 3(\hat{\bm p}\cdot (\bm r-\bm r_s))^2 \left(\frac{1}{{r_{-}^{5/2}}}+\frac{1}{r_{+}^{5/2}}\right)\right)\hat{x} \\ 
&+(y-y_s)\left( \left(\frac{-1}{r_{-}^{3/2}}+\frac{-1}{r_{+}^{3/2}} \right)+ 3(\hat{\bm p}\cdot (\bm r-\bm r_s))^2 \left(\frac{1}{{r_{-}^{5/2}}}+\frac{1}{r_{+}^{5/2}}\right)\right)\hat{y}  \\
&+\left(\frac{(h-z)}{r_{-}^{3/2}}+\frac{(h+z)}{r_{+}^{3/2}} +3(\hat{\bm p}\cdot (\bm r-\bm r_s))^2\left(\frac{h+z}{r_{+}^{5/2}}-\frac{h-z}{r_{-}^{5/2}} \right)\right)\hat{z} \Bigg],   \\ 
\end{split}
\end{equation} 

\noindent where $\hat{\bm p}\cdot (\bm r-\bm r_s)=p_x(x-x_s)+p_y(y-y_s)$, and $r_{\pm} = (x-x_s)^{2} + (y-y_s)^{2} + (z \pm h)^{2}$.

We consider the case where the induced flow is probed by  fluid parcels that are far from the free-slip surface, compared to the position of the swimmers. Therefore, we make the substitution  $z_s=\varepsilon\,z$ with $\varepsilon\ll 1$, and perform a Taylor expansion on the flow exerted by a single swimmer, obtaining
\begin{equation*}
\bm{u}_s\left(\bm{r},\bm{r}_s,\bm{\hat{p}}\right)=\bm{u}_{s,\varepsilon=0}+\frac{\partial \bm{u}_s}{\partial \varepsilon}\bigr\rvert_0\varepsilon+\mathcal{O}\left(\varepsilon^2\right).
\end{equation*}

 Evaluating the velocity field at $\bm{r}=(0,0,z)$, with  $\bm{r}_s=(\rho_s\cos\theta_s,\rho_s\sin\theta_s,h)$ and $\bm{\hat p}=(\cos\phi_s, \sin\phi_s,0)$, the approximated far-field velocity field is,
\begin{subequations}
\label{farfieldflow}
\begin{gather}
   u_{s,\rho_s} = \frac{\kappa\,\rho_{s} [2z^{2} - \rho_{s}^2(1 + 3\cos(2(\theta_{s} - \phi_{s})))]}{(z^{2} + \rho_{s}^2)^{5/2}}(\cos \theta_{s},\sin \theta_{s}),\\
   u_{s,z} = \frac{z\kappa[-2z^{2} + \rho_{s}(1+3\cos(2(\theta_{s} - \phi_{s})))]}{(z^{2} + \rho_{s}^2)^{5/2}}.
\end{gather}    
\end{subequations}
Our AC is characterised by a uniform number density equal to $n$ organisms per unit area, with uniformly distributed orientations and positions, so the AC probability density function is $f(\bm{r}_s,\bm{\hat{p}})=n/2\pi$.
The ensemble-averaged flow field induced by an active carpet is $\langle \bm{u} \rangle = \int_{0}^{\infty} \int_{-\pi}^{\pi} \int_{-\pi}^{\pi}f\,\bm{u}_s(\bm{r},\bm{r}_s,\hat{\bm{p}}_s)\,{\rm d}\phi_s\,{\rm d}\theta_s  \,\rho_s{\rm d}\rho_s  =0.$ 
Thus, the mean flow at any instant is equal to zero for an infinite uniform AC, as expected by symmetry. 
However, the variances of the hydrodynamic fluctuations are not. We can derive them as follows:
\begin{subequations}
\begin{gather}
\langle u_i u_j\rangle= \int_{0}^{\infty} \int_{-\pi}^{\pi} \int_{-\pi}^{\pi}\frac{n}{2\pi}\,u_{s,i} u_{s,j}\, {\rm d}\phi_s\,{\rm d}\theta_s  \,\rho_s{\rm d}\rho_s, \label{variances_a}\\
\langle u_x^2\rangle= \langle u_y^2\rangle=\frac{11\pi n\kappa^2}{16z^2},\hspace{0.25cm} \langle u_z^2\rangle=\frac{9\pi n\kappa^2}{8z^2}. \label{variances_b}
\end{gather} 
\label{variances}
\end{subequations}
Thus, the intensity of the hydrodynamic fluctuations is proportional to the surface density of microswimmers, $n$.
Moreover, it scales with the square of the individual flow strength, $\kappa^2$, and is the same for swimmers of the puller and pusher type. Finally, the fluctuations scale with the inverse of the distance squared, $1/z^2$, so they are stronger near the AC and weaker farther away. 

To test our analytical theory, we perform numerical simulations where the ACs are implemented explicitly as a large set of discrete, non-interacting point dipoles. They are randomly distributed on a square horizontal surface with a side length of $L=Nh$,  where $N$ is a large integer that varies from $10^3$ to $10^4$. The surface density of the microswimmers within the carpet is fixed to $n=0.1/h^2$, so the total number of organisms in the carpet is given by $N_s= n L^2 = 0.1 N^{2}$. 

To evaluate the variances of the hydrodynamic fluctuations shown in equation~\eqref{variances_b}, we compute the average $\langle \cdot \rangle$ over a large ensemble of independent AC configurations. 
In each AC configuration, $N_s$ microswimmers are given new random positions and random orientations within the $x$-$y$ plane where the colony lives. 

In what follows, we choose characteristic units that are natural for \textit{Escherichia coli} bacteria, i.e.
time and length units are fixed such that $h=1$~$\rm \mu m$ and $\kappa=30$~$\rm \mu m^3\,s^{-1}$ \citep{drescher2011fluid}.

We then consider $N_t$ neutrally buoyant and non-Brownian particles to gather statistics for the stochastic tracer dynamics. The equation of motion of the $\alpha$-th tracer is controlled by the flows induced by the AC:
\begin{equation}\label{single_tracer_eq}
\frac{{\rm d}\bm{r}_\alpha}{{\rm d}t}= \sum_s^{N_s}\bm{u}_s(\bm{r}_\alpha,\bm{r}_s,{\bm{\hat {p}})},
\end{equation} 
where $\bm{u}_s(\bm{r}_\alpha,\bm{r}_s,\bm{\hat{p}})$ is the velocity field produced by an individual microswimmer at a position $\bm{r}_s$ and orientation $\bm{\hat{p}}$, acting on a tracer placed at $\bm{r}_\alpha$, using either the full expression from equation~\cref{totstresslet} or the far field approximation from equation~\cref{farfieldflow}.

\vspace{-0.5cm}

\section{Results}
\label{sec:Results}
\subsection{Hydrodynamic fluctuations}
\label{sec:Hydrodynamic fluctuations}
We first focus on examining the velocity variance. Numerically, we produce a large ensemble of AC configurations and evaluate each variance component for a given initial tracer depth $z_0$. The simulation box has a size $L=10^3$, in which $N_s=10^5$ swimmers were placed. \Cref{fig:flowfield}$(b)$ compares the analytical solution for the variance found in equation~\eqref{variances_b} (black and dashed lines) with full numerical simulations (markers). Notice that the horizontal fluctuations, $\langle u_x^2\rangle$ and $\langle u_y^2\rangle$, are about 60\% larger than the vertical fluctuations, $\langle u_z^2\rangle$, showing a significant anisotropic behaviour. The results show that the far-field theory is less accurate close to the AC, $z_0\lesssim 10$, but it offers a good approximation at long distances from the AC. 

These results show that the hydrodynamic fluctuations are space dependent in the vertical direction, decaying as $\langle u_i^2\rangle\propto 1/z_0^2$ through the fluid column. Compared to a solid no-slip boundary, where $\langle u_i^2\rangle\propto 1/z_0^4$ \citep{guzman2021active}, these fluctuations near a liquid-air interface are much longer in range, giving rise to a larger region of influence.

In \cref{Appendix_A}, we also consider more detailed models of active carpets that more closely resemble experimental observations \citep{ahmadzadegan2019hydrodynamic, li2014hydrodynamic, bianchi2017holographic}. 
Instead of modelling the AC as a perfectly planar thin 2D sheet with swimmers that are aligned parallel to the surface, we also consider the cases of a thick AC, a thin AC of swimmers with non-zero orientation angles, and the two combined cases. In all cases, we observe that the variances do not change significantly compared to the initial case, so henceforth we continue with that AC description.

\subsection{Single-particle diffusivity}
Secondly, we examine the diffusion of particles that is caused by the hydrodynamic fluctuations generated by the AC. To explicitly obtain their diffusivity, we consider the single-particle mean-squared displacement (MSD), given by
\begin{equation*}
\langle \Delta r_{i}\Delta r_{j} \rangle 
= \sum_{t' = 0}^{\Tilde{n}\tau}\sum_{t'' = 0}^{\Tilde{n}\tau} \Big\langle u_{i}(\bm{r}_{0}, t')u_{j}(\bm{r}_{0}, t'') \Big\rangle \Delta t' \Delta t'',
\end{equation*}
where $\Tilde{n}$ is the total number of time steps in the simulations and $\tau$ is selected such that matches the characteristic time of swimmers reorientation. We consider the case where the flow fields are uncorrelated between two consecutive time steps. Each time step corresponds to a different independent snapshot of the AC with microswimmers that have randomly sampled positions and orientations. 
Then we have the relation $\langle u_{i}(\bm{r}_{0}, t')u_{j}(\bm{r}_{0}, t'') \rangle = \langle u_{i}(\bm{r}_{0})u_{j}(\bm{r}_{0}) \rangle \delta_{t't''} $, leading to the following expression for the MSD: 
\begin{equation*}
\langle \Delta r_{i}\Delta r_{j} \rangle = \langle u_{i}(\bm{r}_{0})u_{j}(\bm{r}_{0})\rangle \sum_{t' = 0}^{\Tilde{n}\tau}\sum_{t'' = 0}^{\Tilde{n}\tau} \delta_{t't''}\Delta t'\Delta t'' = \langle u_iu_j \rangle\Tilde{n}\tau^{2} = \langle u_iu_j \rangle\tau\,t_{f},
\end{equation*}
where $t_{f} = \Tilde{n}\tau$ denotes the final integration time. Therefore, we have 
$\langle \Delta r_{i}^{2}\rangle = \langle u_{i}^{2} \rangle \tau\,t_{f} \equiv 2D_{i}^{s}t_{f}$,
so the single-particle diffusivity is
\begin{equation}
     D_{i}^{s} \equiv \frac{1}{2}\langle u_{i}^{2} \rangle \tau.
\label{FD_MSD}
\end{equation}
To verify this result numerically, we place tracers initially at a depth $z_0$ and we compute their MSD for $\Tilde{n}=10^2$ time steps. \Cref{fig:flowfield}({\it c}) shows the single-particle diffusivity $D_{i}^{s}$ for $i=z$ as a function of $z_{0}$ obtained from the simulations and from the theory. The simulated observable $\langle u_z^{2} \rangle \tau/2$ is shown as green squares, and the diffusivity, $D_{z}^{s}$, obtained from the simulated MSDs, is displayed as blue circles. These markers collapse onto a single curve, which is compared with the theory from equation~\eqref{FD_MSD} shown by the orange line. Indeed, the results exhibit strong agreement between theory and simulations. 
As predicted, the diffusivity decays with an exponent of $-2$ for large $z_0$. 
This implies that particle stirring significantly intensifies near the AC, potentially enhancing particle encounters, which we discuss next. 

\subsection{Velocity pair correlation}
\label{sec:vel_corr}
Thirdly, we look at how pairs of suspended particles respond to fluid distortions caused by the AC. Although the generated flow is time-uncorrelated, the spatial structure is more complex: Particles that are initially located close to each other follow adjacent streamlines that are almost parallel to each other, so their motions remain correlated at short distances. By contrast, with larger initial separations, the particles follow different streamlines and separate further from each other. 
This relative motion is quantified by the velocity pair correlation function, 
\begin{equation}
\label{pair_corr_def}
    g_{i,p}(d_0, z_0) = \frac{\langle u_{i}(\bm{r}_1) u_{i}(\bm{r}_2)\rangle }{\sqrt{\langle u_{i}^{2}(\bm{r}_1)\rangle \langle u_{i}^{2}(\bm{r}_2) \rangle}},
\end{equation}
where the initial positions of the two particles are $\bm{r}_1 = (d_0/2\sqrt{2}, d_0/2\sqrt{2}, z_0)$ and $\bm{r}_2 = (-d_0/2\sqrt{2}, -d_0/2\sqrt{2}, z_0)$, which are located at the same distance from the carpet $z_0$ and separated laterally by $d_0$.
Notice that the velocity pair correlation is symmetric in the $x$ and $y$ directions. 

\begin{figure}
\centerline{\includegraphics[width=1\linewidth]{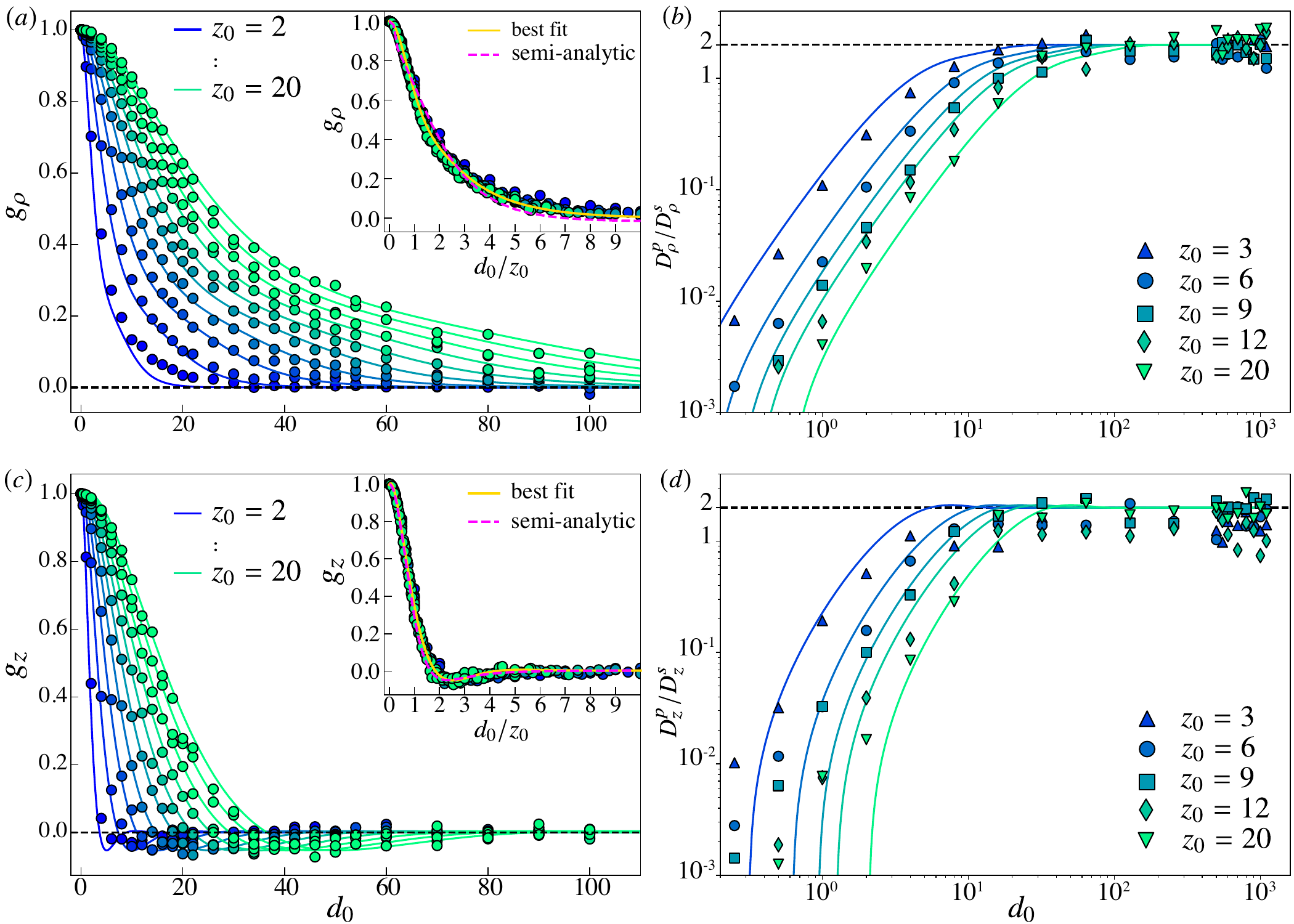}}
\captionsetup{width=1\linewidth}
\caption{($a$) Velocity pair correlation in the planar coordinate, $g_{\rho}$, as a function of horizontal particle separation. Coloured circles represent simulation results for different depths, ranging from $z_0 = 2$ (blue) to $z_0 = 20$ (light green), each separated by $\Delta z_0=2$. Lines represent the best fit to the model from equation ~\eqref{gp_fit}. Inset: These curves collapse when plotted as $d_0/z_0$. The yellow line is the best fit from equation ~\eqref{gp_fit}, and the pink dashed line is the semi-analytical result obtained by numerically integrating $g_{\rho}(d_0, z_0)$. 
($b$) Pair diffusivity in the planar coordinate, $D^{p}_{\rho}$, as a function of horizontal particle separation, normalised by the single-particle diffusivity. Markers represent simulation values for different depths, $z_0 $. The black dashed line is the theoretical asymptotic value from $D_{i}^{p\infty}$. ($c$) and ($d$) show the pair correlation and the pair diffusion in the vertical coordinate, $g_z$ and $D^p_{z}$, respectively.}
\label{fig:Pair_results}
\end{figure}

Numerically, we sample over $10^{4}$ ensembles, each with $N_s=10^7$ swimmers, to measure the velocity using \eqref{totstresslet} at the positions $\bm{r}_1$ and $\bm{r}_2$ given above, with $z_0$ in the range $[2,20]$ and  $d_0$ in the range [$10^{-3}$, $10^3$]. The resulting velocity pair correlations, in the vertical and horizontal orientations, are shown as circles in \cref{fig:Pair_results}($a,c$). As predicted, we find that pairs become progressively more uncorrelated with distance $d_0$. Moreover, the separation required for the pairs to become uncorrelated grows with $z_0$. Notably, when the velocity pair correlations are plotted in terms of the scaled distance $d_0/z_0$, there is a collapse for all depths onto a single curve, as shown in the insets of \cref{fig:Pair_results}($a,c$). Globally, velocities decorrelate at different rates, depending on the velocity direction, with particles decorrelating at distances larger than $d_0\gtrsim 7 z_0$ in the horizontal direction and $d_0\gtrsim 4 z_0$ in the vertical direction.
  
First, we propose a heuristic model for the collapsed velocity correlation of the form
\begin{equation}
\label{gp_fit}
    g_{i,p}^{\text{model}}(\xi) = \exp(-\xi)[A_1 \sin(\xi) + \cos(\xi)] + \exp(-\xi/A_2)[A_3 + A_4\xi],
\end{equation}
where $\xi = d_0/z_0$. Using least squares to fit this model to our simulation data, we find that the optimal fitting parameters are  in the horizontal direction $A_1 = 0.097$, $A_2 = 1.3$, $A_3 = -0.014$, $A_4 = 0.89$ , and in the vertical direction $A_1 = -0.18$, $A_2 = 0.52$, $A_3 = -0.09$, $A_4 = 1.35$. The resulting model is shown as coloured lines for the different $z_0$ values in \cref{fig:Pair_results}($a,c$), and also in the inset. 

Analytical progress can be made by using the far-field approximation, where $g_{i,p}$ is obtained by evaluating the velocity from equation~\eqref{farfieldflow} and the variances from equation~\eqref{variances} at the positions $\bm{r}_{1}$ and $\bm{r}_{2}$, 
\begin{equation}
\label{pair_corr_integral}
    g_{i,p}(d_0, z_0) = \frac{1}{\langle u_i^{2} \rangle}\int_{0}^{2\pi}\int_{0}^{2\pi}\int_{0}^{\infty} u_{i}(\bm{r}_1) u_{i}(\bm{r}_2)\frac{n}{2\pi} \rho_{s}{\rm d}\rho_{s}{\rm d}\theta_{s}{\rm d}\phi_{s}.
\end{equation}
It is challenging to find a closed-form expression for the above integral, but it can be evaluated numerically; the semi-analytical result is shown in the inset of \cref{fig:Pair_results}$(a)$.
The theoretical result, which is in strong agreement with our simulations, underscores the crucial role of the vertical distance to the AC in our study:
It does not only determine the strength of the hydrodynamic fluctuations but also the relative motion of the flow structures, with strong correlation at larger distances for flows parallel to the AC and fast decaying correlations in the vertical direction, which is also evident with the heuristic model in equation~\eqref{gp_fit} with the parameter $A_2$ representing the rate at which particles decorrelate with distance, where $A_2^{\rho}=1.3$ and $A_2^{z}=0.52$. 
Particles at greater depths tend to move together synchronously, which directly impacts their pair diffusivity, as we show next.

\subsection{Pair diffusion}
\label{sec:PairDiffusion}
We now explore how an AC drives can mix two particles up, as a function of their distance. 
This is quantified by the `pair diffusivity'. Numerically, we sample the carpet as described previously. Two tracers are put initially at the positions $\bm{r}_1$ and $\bm{r}_2$, after which they follow the equation of motion, equation ~\eqref{single_tracer_eq}. The integration time step is $\Delta t = 10^{-3}$. Averages are performed over $300$ pair trajectories.
The pair diffusivity $D_{i}^p$ is obtained by measuring the pair's distance squared, $\Delta\rho^2(t)=\left(|\bm{r}_1(t)-\bm{r}_2(t)|-d_0\right)^2$.
 After the simulation ends, each $\Delta\rho^2$ curve is fitted with a power-law function of the form $4D_{i}^p\,t^{b}$. Results show that all curves follow a diffusive regime ($b \sim 1$).  
 
\Cref{fig:Pair_results}({\it b})({\it d}) displays the resulting pair diffusivity $D_{i}^p$, normalised by the single-particle diffusivity $D^{s}_{i}$, as a function of the initial horizontal particle separation $d_0$. 
 For small separations, the particles follow almost the same streamlines, so their relative motion and pair diffusivity is small. Indeed, it is difficult to mix particles that are close to each other, as we observe in the kitchen when stirring macaron batter \citep{Mathijssen2023}. At large particle separations, it gets easier to mix them, so the pair diffusivity increases with $d_0$ until it plateaus. 
 
To predict this analytically, we define $ \bm{r}_{12}= \bm{r}_{1}- \bm{r}_{2}$. Similar to the single-particle MSD calculation, the pair MSD is then computed as follows,
\begin{equation}\label{pair-MSD}
    \langle \Delta r_{12i} \Delta r_{12j} \rangle = \langle \Delta r_{1i}\Delta r_{1j} \rangle + \langle \Delta r_{2i}\Delta r_{2j}\rangle - \langle  \Delta r_{1i}\Delta r_{2j} \rangle - \langle  \Delta r_{2i}\Delta r_{1j} \rangle,
\end{equation}
where $\Delta r_{\alpha i}$ denotes the displacement of the particle $\alpha$ along coordinate $i$ and analogously for $\Delta r_{12 i}$. On average, terms 1 and 2 on the RHS of equation ~\eqref{pair-MSD} are the same, and terms 3 and 4 are also equal to each other. Therefore, 
    $\langle \Delta r_{12i} \Delta r_{12j}\rangle = 2\langle \Delta r_{1i}\Delta r_{1j}\rangle - 2\langle  \Delta r_{1i}\Delta r_{2j} \rangle$.
Considering $i = j$, the first term is $\langle \Delta r_{i}\Delta r_{i} \rangle = 2 D_{i}^{s} t$, with $D_i^{s}$ the single diffusivity in the $i$-coordinate. The second term can be estimated in a similar way as in the single MSD calculation, $\langle \Delta r_{1i}\Delta r_{2i} \rangle = \sum_{t' = 0}^{\Tilde{n}\tau}\sum_{t'' = 0}^{\Tilde{n} \tau} \langle u_{1i}u_{2i} \rangle \Delta t'\Delta t''$ (with $\tau=\Delta t$). The function to be integrated is the complete spatiotemporal pair correlation velocity, $\langle u_{1i}u_{2i} \rangle(d_0, z_0, t) = \langle u_i^{2} \rangle g_{i,p}(d_0, z_0) \delta_{t't''}$ where $g_{i,p}$ is the spatial pair velocity correlation defined in \cref{pair_corr_def}. With this, we obtain 
$\langle \Delta r_{1i}\Delta r_{2i} \rangle = \langle u_{i}^{2} \rangle g_{i,p} \sum_{t'=0}^{\Tilde{n}\tau}\sum_{t''=0}^{\Tilde{n}\tau} \delta_{t't''} \Delta t' \Delta t'' = \langle u_{i}^{2} \rangle g_{i,p} \tau\, t_{f}$.    
Thus,
$\langle (\Delta r_{12i})(\Delta r_{12j})\rangle = 2\left(2D_{i}^{s} - \langle u_{i}^{2} \rangle \tau g_{i,p}\right)t_{f}$, where we obtain the pair diffusivity
\begin{equation}
D_{i}^{p} \equiv 2D_{i}^{s} - \langle u_{i}^{2} \rangle \tau g_{i,p} = 2D_{i}^{s}(1 - g_{i,p}).
\label{D_pair_def}
\end{equation}
This expression depends upon the variance and single diffusivity relation in equation~\eqref{FD_MSD}. Since the function $g_{i,p}(d_0, z_0)$ decays exponentially with the distance $d_0$, the asymptotic pair diffusivity is readily predicted for large $d_0$, 
\begin{equation}
    \lim_{d_0 \to \infty} D_{i}^{p} \equiv D_{i}^{p\infty} = 2D_{i}^{s}.
\label{PairDiffusionAsymptote}
\end{equation}
\Cref{fig:Pair_results}({\it b})({\it d}) shows this theoretical result for the pair diffusivity \eqref{D_pair_def}, using the $g_{i,p}^{\text{model}}$ in \eqref{gp_fit}. This result agrees well with the simulations, as shown for various distances $z_0$.

Interestingly, the observed diffusive behavior in our study bears a resemblance to turbulent systems, as seen in \cite{belan2019pair}, where a pair of tracers within a 3D bath of microswimmers also exhibited an asymptotic diffusivity twice that of their self-diffusion, reaching numerical values of $D^p_{\infty}\approx2\mu$m$^2$/s which is in the same order of magnitude in our case when we choose $\tau =\tau_R=10$ as a typical memory time for \textit{Escherichia coli} bacteria and, $ n = 0.01,\; \kappa = 30,\; z_0 = 10,\; d_0 = 10$, we get $D_{\rho}^p\approx1.39\mu$m$^2$/s, with horizontal and vertical self-diffusivities $D^s_{\rho} \approx 1.94 ~\mu\text{m}^2/\text{s}, D^s_z \approx 3.18 ~\mu\text{m}^2/\text{s}$, respectively which are one order of magnitude larger than thermal diffusion for a passive spherical micronsized particle $D_{T} \approx 0.22 ~\mu\text{m}^2/\text{s}$.

To put these numbers in perspective, in natural aquatic environments, bioconvection propelled by Chromatium Okenni (with a body length of 1-4 $\mu$m) has been demonstrated to notably boost the actual diffusion of heat within their ecological niche \citep{sommer2017bacteria}. This enhancement, observed to be at least an order of magnitude, was documented in `Lago di Cadagno', Switzerland \citep{sepulveda2019convection}. Similar diffusion enhancement has been measured in laboratory experiments driven by E.~coli \citep{singh2021bacterial}. Remarkably, the thermal diffusion augmentation resulting from the collective swimming of microorganisms is not significantly different from the enhancement in vertical heat diffusion around the thermocline induced by wind-driven shear flows in strongly stratified lakes, as evidenced in Lac Léman \citep{fernandez2021seasonality,sepulveda2023spatial}.

Our results reveal that the distance parameter $z_0$ significantly affects diffusivity. The mixing becomes more intense and converges to the asymptotic value when particles are closer to the AC. This phenomenon can be attributed to amplified hydrodynamic fluctuations near the carpet, which disrupt spatial correlations in fluid flows. The parameter $d_0$ is also crucial; closer particles exhibit lower diffusion from each other compared to those at greater distances, resulting in synchronized movement and more similar trajectories, with vertical pair diffusion reaching the asymptotic value faster than horizontal pair diffusion as expected from the pair velocity correlation behavior. Consequently, the pair diffusivity might offer fundamental insights into aggregation phenomena, which we study next.

\subsection{Particle aggregation}\label{sec:ParticleAggregation}
We investigate whether fluctuations caused by ACs can initiate the aggregation of spherical particles of finite radius $r_{0}$. 
To model this, we consider a short-ranged pairwise sticky force between the particles. We consider the Morse model, given by 
\begin{equation*}\label{eq:Morse_model}
\bm{F}_{M}(\bm r) = 2UW\left[1 - e^{-W(r-r_{\text{eq}})} \right]e^{-W(r - r_{\text{eq}})}\hat{\bm{r}}, \end{equation*}
where $r=|\bm{r}|$ is the distance between the tracers. Here, $r_{\text{eq}}$ is the pair equilibrium distance, while $U$ and $W$ denote the depth and width of the potential well, respectively. The force is scaled with the Stokes mobility, so it has units of velocity. By fixing $r_\text{eq}$ to twice the radius $r_{0}$, $U$ to $10^{-9}$, and $W$ to 30, we set the potential minima along $r$, allowing for the precise adjustment of equilibrium distances for tracers with different radii $r_{0}$.

Initially, at $t=0$, the particles are placed in a square lattice of $N_t = 19\times19 = 361$ particles located at the same depth $z_0$. The distance between their centres is equal to $d_{0} = \Tilde{d}_0 + 2r_{0}$, where $\Tilde{d}_0$ is their border-to-border distance.
We always set $\Tilde{d}_0 = 0.8$ and vary $d_0$ or equivalently $r_{0}$. For $t>0$, the particles start moving in the $x$-$y$ plane at a fixed depth $z_0$, obeying the equation of motion \eqref{single_tracer_eq}, with the addition of the inter-particle sticking forces that allows them to collide and aggregate to each other $ \frac{{\rm d}\mathbf{r}_\alpha}{{\rm d}t} = \sum_s^{N_{\text{s}}}\mathbf{u}_s(\mathbf{r}_\alpha,\mathbf{r}_s,\mathbf{\hat{p}})+\sum_{\alpha\neq \beta}^{N_{\text{t}}} \mathbf{F}_M^{\alpha\beta}(\mathbf{r}_\alpha, \mathbf{r}_\beta, t),$.

\begin{figure}
\centerline{\includegraphics[width=1\linewidth]{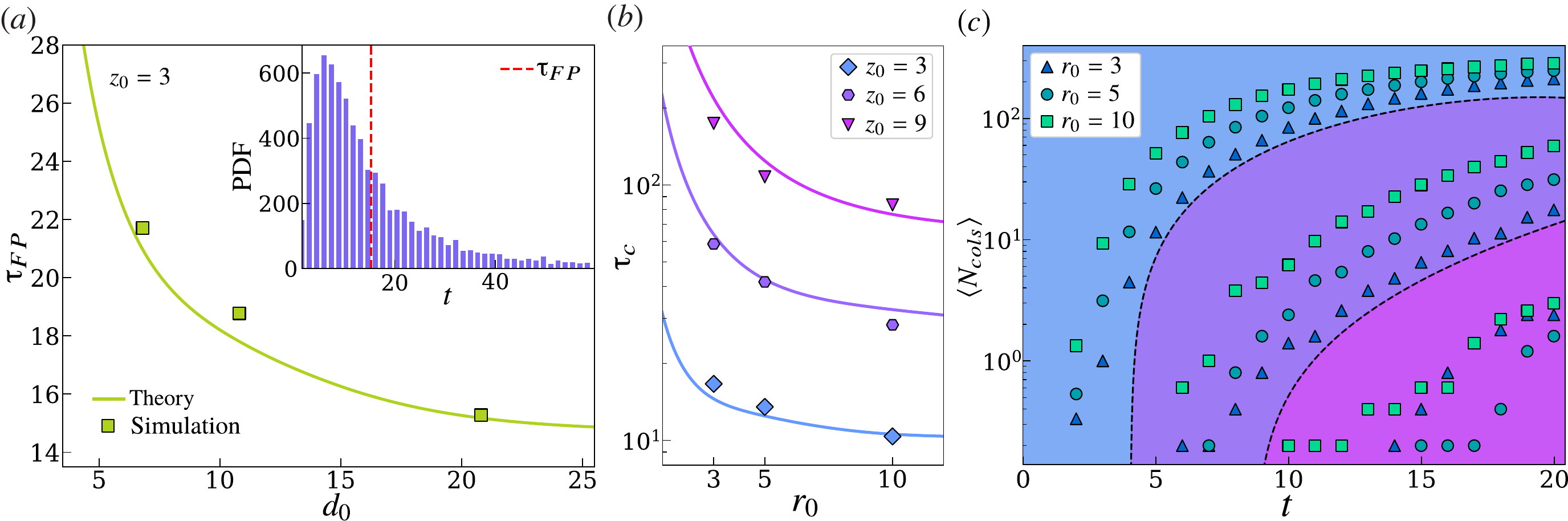}}
\captionsetup{width=1\linewidth}
\caption{($a$) Mean first passage time of particle collisions as a function of the horizontal particle separation, $d_0$, for particles initially located at $z_0 = 3$. Squares represent simulations and the line is the prediction $\uptau_{\text{FP}} \approx \Tilde{d_0}^{2}/4 D_{\rho}^p.$ Insite: Histogram of first passage times. The vertical red dashed line is the mean first passage time. ($b$) Clearance time against particle radius for three depths. Symbols show simulation results, and solid lines show $\uptau_c\sim0.7\uptau_\text{FP}$. ($c$) Average accumulated number of collisions over time. Markers show averaged simulation results for different particle radii, ranging from $a = 3$ (triangles) to $a = 10$ (squares). The background colours represent different depths, ranging from $z_0 = 3$ (light blue) to $z_0 = 9$ (pink).}
\label{fig:aggregation}
\end{figure}

We investigate the particles aggregation by examining the histogram of first passage times, defined as the time taken for particles to collide. Results are shown in the inset of \cref{fig:aggregation}({\it a}); the red dashed line shows the mean first passage time, $\tau_{\text{FP}}$. This observable can also be estimated analytically by equating the pair MSD in \eqref{D_pair_def} to the border-to-border distance between particles, $\Tilde{d_0}^{2} \approx \langle \Delta r_{1} \Delta r_{2} \rangle = 4 D_{\rho}^p\tau_{\text{FP}}$, so the mean first passage time is
\begin{equation}
\tau_{\text{FP}} \approx \frac{\Tilde{d_0}^{2}}{4 D_{\rho}^p}.
\label{TFP_prediction}
\end{equation}
The theoretical prediction for $\uptau_{FP}$ agrees well with the numerical simulations, as shown for $z_0 = 3$ by the solid green line in \cref{fig:aggregation}($a$). Therefore, $\uptau_\text{FP}$ provides a reliable estimate for the average collision time between two particles due to the hydrodynamic stirring induced by the AC.
    
In addition, we calculate the average `clearance time' denoted as $\uptau_c$. This is the time taken for half of the particles in the suspension to collide, representing a fundamental timescale indicator for aggregation and cluster formation \citep{font2019collective, wang1998collision}. The graph shown in \cref{fig:aggregation}({\it b}) illustrates the clearance time as a function of particles radius $r_{0}$ and depth $z_{0}$, where we assume that particles stick together after a collision. Our results show that, as the size of the particle $r_{0}$ increases, the clearance time decreases. Moreover, $\uptau_{FP}$ and $\uptau_{c}$ are proportional to each other since they both arise from the same physical process. Empirically, we observe that $\uptau_c \sim 0.7 \uptau_{\text{FP}}$ for every $z_0$, as shown by the lines in \cref{fig:aggregation}({\it b}), which agrees well with the simulations.

Last, we measure the time evolution of the average accumulated number of collisions $\langle N_{cols}\rangle$ for this configuration. \Cref{fig:aggregation}$(c)$ shows $\langle N_{cols}\rangle$ over time for each depth $z_0$. The results show a close correlation between $\langle N_{cols}\rangle$ and $\uptau_c$. Collisions become more frequent as we move closer to the AC and increase with larger particle radius. The latter suggests that the pair correlation function $g_{\rho}$ in \eqref{pair_corr_def} governs both the collisions and the measured aggregation timescale. As the particle size increases, the correlation among their centres decreases, making it easier for them to follow different trajectories and ultimately leading to more collisions. Likewise, the intense fluctuations near ACs disrupt this correlation, increasing the likelihood of collisions.

\section{Concluding remarks}
\label{sec:Conclusion}
Here, we have established analytical connections between the flows generated by active carpets and the aggregation dynamics of suspended particles. For this, we have investigated broadly used fundamental metrics that characterised suspended particle dynamics: the single and pair diffusivities and particle aggregation. Our key findings include:
(i) An analytical solution for hydrodynamic fluctuations produced by \textit{active carpets} near fluid-air interfaces;
(ii) the emergence of space-dependent and anisotropic diffusion, which decreases quadratically with distance;
(iii) the role of hydrodynamic fluctuations in facilitating pairing encounters, where particle aggregation is favoured for large compared to small particles;
(iv) the mean first-passage time between collisions decreases as particles move farther apart;
(v) in the close vicinity of active carpets, intense hydrodynamic stirring accelerates clearance times and particle aggregation processes. 

Although our study focused on active carpets, the employed methodology can be applied to other swimmer configurations. In particular, \eqref{D_pair_def}, which relates the pair diffusivity to the velocity pair correlation, is a valuable tool to to study dynamical aggregation driven by active fluctuations.
This research highlights the pivotal role of biologically driven flows in the transport and spatial organisation of particles in aquatic systems, serving as a noteworthy example of an out-of-equilibrium system that remains analytically tractable.\\

\noindent\textbf{Acknowledgments}. We are grateful for the helpful feedback given Hartmut L{\"o}wen and Felipe Barros. We also thank the anonymous reviewers for their constructive feedback.\\

\noindent\textbf{Fundings}.  G.A., R.S. and F.G.-L. have received support from the ANID – Millennium Science Initiative Program – NCN19 170, Chile. F.G.-L. was supported by Fondecyt Iniciación No.\ 11220683. H.N.U. and A.J.T.M.M. were supported by start-up grants from the University of Pennsylvania.
A.J.T.M.M. acknowledges funding from the United States Department of Agriculture (USDA-NIFA AFRI grants 2020-67017-30776 and 2020-67015-32330), the Charles E. Kaufman Foundation (Early Investigator Research Award KA2022-129523) and the University of Pennsylvania (University Research Foundation Grant and Klein Family Social Justice Award).\\

\noindent\textbf{Declaration of interests}. The authors report no conflict of interest.
\vspace{-0.25cm}

\section*{Author ORCIDs}
Gabriel Aguayo ~~ \href{https://orcid.org/0009-0004-4915-2707}{https://orcid.org/0009-0004-4915-2707}\noindent\\
Arnold Mathijssen ~~ \href{https://orcid.org/0000-0002-9577-8928}{https://orcid.org/0000-0002-9577-8928}\noindent\\
Hugo N. Ulloa ~~ \href{https://orcid.org/0000-0002-1995-6630}{https://orcid.org/0000-0002-1995-6630}\noindent\\
Rodrigo Soto ~~\href{https://orcid.org/0000-0003-1315-5872}{https://orcid.org/0000-0003-1315-5872}\noindent\\
Francisca Guzman-Lastra ~~ \href{https://orcid.org/0000-0002-1906-9222}{https://orcid.org/0000-0002-1906-9222}\noindent


\appendix
\section{Thick and tilted active carpet}\label{Appendix_A}

In nature and laboratory experiments, swimming microorganisms form spatial distributions that are more complex than those of a single monolayer. They can create `thick' films where the number density of microswimmers decays with depth. Also, it has been observed that microswimmers do not always orient parallel to air-liquid interfaces; observations also show that, in some cases, their average angle orientation is slightly out-of-the-plane, which we denote as `tilted' swimmers \citep{ahmadzadegan2019hydrodynamic, li2014hydrodynamic, bianchi2017holographic}.
We extended our results to three different cases: a thick carpet, a tilted carpet, and a thick tilted carpet.
Hence, we performed simulations to quantify the changes in the hydrodynamic fluctuations.

\begin{figure}
\centerline{\includegraphics[width=1\linewidth]{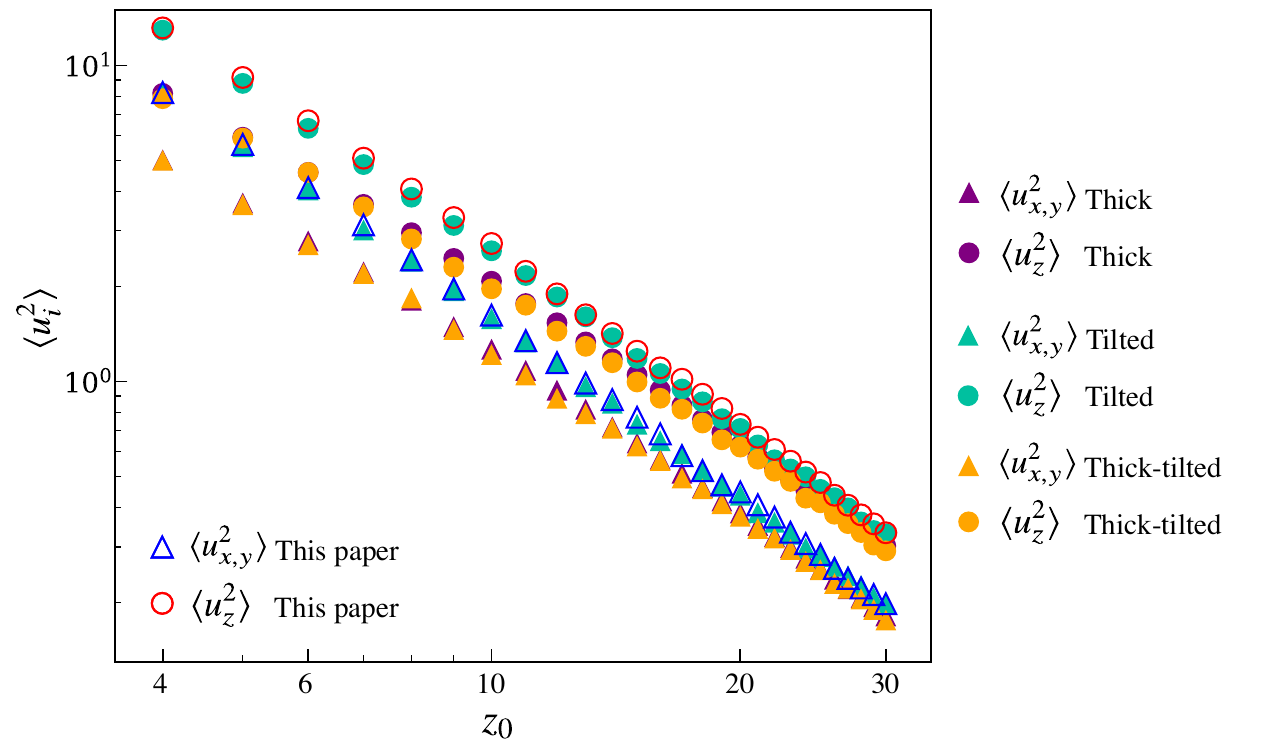}}
\captionsetup{width=1\linewidth}
\caption{Comparison between variances measured when the AC is thick and have planar swimmers (purple markers), when the AC is flat and have tilted swimmers (light-green markers), when the AC is thick and have tilted swimmers (orange markers) and when the AC is flat and have planar swimmers (hollow markers, the manuscript case).}
\label{fig:apendix}
\end{figure}

For the thick carpet, we consider swimmers that do not form a planar 2D sheet, but they are distributed in 3D according to a vertical exponential function,
\begin{equation}
    f(z_s) = 2\exp(-2(z_s-h)),
\end{equation}
where $z_s$ is the swimmer's vertical position.
Using this distribution, in practice,  swimmers are found up to approximately $z_s\sim 5h$. In this scenario, we defined the `average carpet position' at $\Bar{z} = 2.5h$. With this, we computed the variances for different depths in the same way as in the manuscript. \Cref{fig:apendix} displays these results in purple symbols. We notice that the thick carpet variances conserve the thin AC behaviour with vertical fluctuations larger than planar fluctuations decaying with the same power-law. We also observe, quantitatively, some changes close to the active carpet. In contrast, for far distances, the values are alike. 

For the tilted case, we consider microswimmers with vertical out-of-plane angles from $-10^{\circ}$ to $10^{\circ}$, distributed according to a Gaussian with zero mean and a standard deviation of $5^{\circ}$.
\Cref{fig:apendix} displays the results in green symbols. Coloured and hollow markers match perfectly well; therefore, the $z_0$ dependency holds. Only a slight decrease in magnitude is observed in the tilted case. This change could be understood as a geometric factor (a little less than the unity) multiplying the planar AC used in this manuscript.
In addition to the first case, we also performed numerical simulations using a thick carpet containing tilted swimmers (both features together) to ensure completeness. In \cref{fig:apendix} with orange markers, it can be seen that this case is almost the same as the thick-carpet-with-planar-swimmers case. Hence, we can conclude that a thick carpet affects hydrodynamic fluctuations more than a carpet with tilted swimmers. In all cases, however, the effect is small and does not change the order of magnitude of the variance. Therefore, such variants in the AC properties do not obscure the most fundamental results and concluding remarks presented in the manuscript; moreover, they do not break the anisotropic behaviour of fluctuations, highlighting the universality of our results.

\bibliographystyle{jfm}
\bibliography{jfm_R1}

\end{document}